# Mechanical activation of reversible bonds by low amplitude high frequencies excitations

Maziar Heidari [1]†, Théophile Gaichies[1], Ludwik Leibler[1], Matthieu Labousse[1]*,

[1]Gulliver, CNRS, ESPCI Paris, Université PSL, 75005 Paris, France

†Present address: Theoretical Biophysics Department, Max Planck Institute for Biophysics, 60438 Frankfurt am Main, Germany

*Corresponding author. Email: matthieu.labousse@espci.psl.eu

Abstract: Reversible covalent or supramolecular bonds play an important role in materials science and in biological systems. The equilibrium between open and closed bonds and the association rate can be controlled thermally, chemically, by mechanical pulling, ultrasound or catalysts. In practice, these intrinsic equilibrium methods either suffer from a limited range of tunability or may damage the system. Here, we present a non-equilibrium strategy that exploits the dissipative properties of the system to control and change the dynamic properties of sacrificial and reversible networks. We show theoretically and numerically how high-frequency mechanical oscillations of very low amplitude can open or close bonds. This mechanism indicates how reversible bonds could alleviate mechanical fatigue of materials especially at low temperatures where they are fragile. In another area, it suggests that the system can be actively modified by the application of ultrasound to induce gel-fluid transitions and to activate or deactivate adhesion properties.

One-Sentence Summary: We show how mechanical oscillations of low amplitude and high frequency, through a process termed mechanical pumping, can control ligand receptor bonding.

**Introduction** The control of ligand-receptor binding dynamics emerges as a central issue in the adhesives (1), tires and plastics industry. In particular, it is an essential feature for determining the properties of all materials based on reversible networks (2,3) such as supramolecular architectures (4), dynamics polymer network (5), and self-healing systems (6) and for controlling their fracture (7,8) and fatigue resistance (9). Similarly in the biological system, the binding kinetics of receptor-ligand complex plays a crucial role in the cellular functioning. For example, the complexes embedded on the cell membrane are under thermal and active fluctuations with a wide range of spatiotemporal spectrum (10-12). A chemical reaction or ligand-receptor bond is governed by thermodynamics or kinetics (13). Under thermodynamical control, the rate of binding ($k_{o \to b}$) over unbinding ($k_{b \to o}$) is expected to satisfy the Kramers' relation (14) $\frac{k_{b \to o}}{k_{o \to b}} = e^{\Delta F / k_B T}$ where, $\Delta F$, is the free energy difference between the open ($o$) state and the bound ($b$) state, $k_B$, the Boltzmann constant and, $T$, the temperature. In contrast, under kinetic control, the Arrhenius rule (14) implies that reaction pathway possessing the smallest activation energy $E_a$ is the fastest with a transition rate scaling as $\sim e^{E_a / k_B T}$. Although fundamentally different, these two relations dictate that a reaction can only be favored by either changing the energy landscape or the temperature (15). While the change of temperature is, by essence, unselective and can only be realized in a very narrow range, the decrease of an activation energy requires usually a catalytic strategy which is reaction-specific and which represent most of the time a scientific *tour de force*. Alternatively, the binding kinetics of a single receptor and its corresponding ligand may be influenced by external mechanical forces and different loading conditions. Constant force or pulling with constant velocity, have been thoroughly studied theoretically (16-26) and experimentally (27-31). Extensile mechanical states are known to promote the breaking of a bond. Also, oscillatory forcing of a ligand receptor association has been theoretically investigated (32-34) with the main *anstaz* following the spirit of a Bell's like theory (16). Experimentally, ultrasound waves were used to generate such strong oscillatory forces in the bulk to control the reaction and dissociate bonds. Spectacular examples in mechanochemistry, like mechanically-activated catalysis (35), and efficient antibiotic release pathways (36-39), were experimentally designed. Also, liquid-gel transition (40) in Rhodium(I)-Iridium(I) (41), naphthylamide-based (42) and diselenide-crosslinked (43) networks were demonstrated. These networks form a class of ultrasound responsive polymer system (44-52). In these cases, the cavitation induced by the mechanical waves is thought to generate a shearing stress sufficient to break the bonds of the gel network. However, the microscopic mechanisms at stake remains unelucidated but should require strong local displacements. In contrast, we propose and predict, here, an alternative non-equilibrium mechanochemical strategy holding even for nanometric displacements which always favor the bond breaking, and whose core idea is sketched in Fig 1A and 1B.

We consider two systems in which a ligand binds to receptors. In the first system, the ligand is a Brownian molecule, while in the second system, the ligand is attached to the end of a grafted polymer chain. We compare this equilibrium situation with that placed under a mechanical wave which sets the receptors into an oscillatory motion. This action is displacement-controlled. In this article, we demonstrate that this particular constrained motion favors the open state (Fig 1c). We show that, even for a very small motion amplitude, the open state is exponentially favored by friction above a characteristic forcing frequency whenever the system would be under extensile or compressive states. We call this out-of-equilibrium protocol *mechanical pumping*, a term inspired from its optical counterpart owing to some unexpected analogous mechanisms.

Theoretical predictions We introduce the mechanism at stake from a theoretical perspective before realizing a numerical proof of concept. As exemplified in Fig 1A, a surface with attached receptors is set into a displacement-controlled oscillatory motion of amplitude $A$ and angular frequency $\omega$. For simplicity, we start our analysis by considering the case of a Brownian particle. When the Brownian particle detaches from the receptor, it does not feel the presence of the wall anymore and we expect that the open time should be almost independent on the wall motion, an assertion we shall confirm further. So, we focus our analysis on the transitions from the bound to the open state. The mean transition time from bound to open state, $T_{b \to o}^{\text{Brownian}}$, specifically the mean escaping time, is given by $T_{b \to o}^{\text{Brownian}} = -\int_0^{+\infty} t \, \dot{P}_s(t) dt$. In this equation, $P_s(t)$ denotes the survival density probability and is related to the time dependent escaping rate, $k(t)$, by the relation $P_s = \exp(-\int_0^t k(t')dt')$ (33). In addition, in the referential comoving with the wall, the overdamped Brownian particle experiences a friction force along the direction of the wall motion, $F_\gamma(t) = \gamma A \omega \cos(\omega t)$ with $\gamma$ the friction coefficient. As a consequence, we infer the form for time dependent escaping rate $k(t) = k_0 \exp(-E_a(t)/k_B T)$ by considering a time-dependent energy barrier $E_a(t) = E_{a,0} - F_\gamma(t)\sigma$ with, $\sigma$, the effective spatial extension of the potential and, $E_{a,0}$, the binding energy at equilibrium. This adiabatic hypothesis is reminiscent of the Bell Ansatz (16) and has been shown to hold for constant and oscillatory forces (19,22). Note that the displacement-controlled motion of the receptors induces a linear frequency dependence in the amplitude of $F_\gamma(t)$, a situation which shall not be encountered with a force-controlled motion. After calculations and following (22,33), we get

$$T_{b \to o}^{\text{Brownian}} = \frac{T_{b \to o}^{\text{Brownian}}(0)}{I_0\left(\frac{\gamma A \omega \sigma}{k_B T}\right)}, \quad (1)$$

with $T_{b \to o}^{\text{Brownian}}(0)$ the mean transition time when the traps are immobile ($\omega = 0, A = 0$), and $I_0$ the modified Bessel function of first kind of order 0. Furthermore, $I_0$ behaves essentially as an exponentially growing function at the leading order so that balancing the denominator term in Eq. 1 with its nominator $T_{b \to o}^{\text{Brownian}}(0) \propto \exp(E_{a,0}/k_B T)$ predicts a crossover angular frequency, $\omega_{\text{Brownian}}^*$ given by $\gamma A \omega_{\text{Brownian}}^* \simeq E_{a,0}/\sigma$. The latter relation is interpreted as a force balance between the driving friction and the characteristic trapping force field. Guided by this physical picture, the critical force balance in the case of the second system, a ligand attached to the end of a grafted polymer, is expected to scale as $z\gamma A\omega^*_{chain} \simeq E_{a,0}/\sigma$ where $z$ is the characteristic number of monomer that frictions when the wall start moving upwards and $\omega^*_{chain}$ the corresponding crossover angular frequency. By recalling that a segment composed of $z$ monomers relaxes local stresses within a Rouse time $\tau_R = \gamma \frac{(za)^2}{3\pi^2 k_B T}$ (53), we note that a segment of $z^*$ monomers can relax in timescale faster $1/\omega$ when $\tau_R(z^*) = \frac{1}{\omega}$, or equivalently if the segment has $z^* = \frac{\sqrt{3}\pi}{a}\left(\frac{k_B T}{\gamma\omega}\right)^{1/2}$ monomers. As a result, for the case of the ligand attached to the end of a grafted polymer, we predict that the master curve collapsing the mean transition time from bound to open state $T^{chain}_{b\to o}/T^{chain}_{b\to o}(0)$ must be a function solely of the product $z\gamma A\omega_{chain} \propto A(\gamma\omega_{chain})^{1/2}$ and yields

$$T^{Chain}_{b\to o} = \frac{T^{Chain}_{b\to o}(0)}{I_0\left(\frac{z(\omega)\gamma A\omega\sigma}{k_B T}\right)}, \quad (2)$$

with $T^{Chain}_{b\to o}(0)$ the mean transition time when the traps are immobile. Both Equations 1 and 2 involve an $I_0$ function in the denominator. As $I_0$ diverges exponentially, we predict that the transition time from bound to open state is significantly reduced at high frequency. Therefore, when the receptors are mechanically pumped, the detachment process is effectively promoted while the attachment event remains unchanged.

Numerical scheme To assess the relevance of our theoretical predictions, we consider in Fig. 1C, a numerical model system composed of a ligand attached to the end of a grafted polymer (see methods section). It is facing a receptor field positioned at a non-permeable wall moving around a mean position $h_0$ with an amplitude $A$ at a frequency $f = \omega/2\pi$. We use the generic coarse-grained Kremer-Grest polymer model (54,55) and unless explicitly stated otherwise, in what follows, lengths are expressed in monomer size, $a$, units, energy in $k_BT$ units, timescales in monomer relaxation time, $\tau = \sqrt{ma^2/k_BT}$, units (see methods section), with $m$ the mass of particle and frequency in $\tau^{-1}$ unit.

Numerical Results We present in Fig. 1D the typical numerical results by simulating the molecular dynamics of the chain systems, for a rest interwall distance $h_0 = 8$ and oscillation amplitude $A = 2$. We find that the probability of binding $P_b$ sharply decreases when the oscillation of the wall exceeds a characteristic frequency $f^* \simeq 10^{-2}$. For frequency $f < f^*$ the probability of binding reaches a plateau corresponding to its value at equilibrium. In sharp contrast, for frequency $f > f^*$ the probability of binding abruptly decreases to zero and we find a strong difference compared with the equilibrium binding probability. We show in Fig. S1 a top view of the cumulative path travelled by the ligand over a fixed and long period of time ($2 \times 10^5\ \tau$) for increasing frequency. When the wall is immobile, the binding is strong and the ligand spends most of its time attached. In fact, it detaches very rarely and when it happens it either rebinds to the same receptor or explores laterally and rebind to another receptors. Rebinding process is controlled by diffusion and density of receptors when the wall oscillates. As the frequency increases, the detaching events are more probable and the ligand spend more time detached from the receptor field. As a consequence, its lateral diffusion is more pronounced and the ligand visits much more receptors. For example, at $f = 0.33$ the ligand visits about 3 times more receptors than at equilibrium for an equivalent period of times.

We now examine the origin of the mechanical pumping and demonstrate that its efficiency lies in the combined effects of a gap opening and diffusion-controlled rebinding. The gap opening is evidenced in Fig. 2A by comparing the axial position probability distribution $P_z$ of the ligand for increasing frequency. At low frequency, $f = 3.3 \times 10^{-3} < f^*$ for which the ligand is mostly bound, $P_z$ is strongly peaked around $z = h_{\min}$ and $z = h_{\max}$ and there is a pronounced gap of probability between these modes. We explain this gap creation by observing that wall speed is faster in the region $z = h_0$ and slower in the region $z = h_{min}$ and $z = h_{max}$. This non-uniform motion implies that the receptor spends more time around $z = h_{min}$ and $z = h_{max}$ which concentrates the position probability distribution $P_z$ at these two positions. We refer to as *pumping state* (*intermediate state*, respectively) the bonding events corresponding to the vicinity of peak at $z = h_{max}$ ($z = h_{min}$, respectively), and to *open state* the unbinding events in $z \in [0, h_{min}]$. At low frequency, ($f = 3 \times 10^{-3}$ in Figs 2A and 2B) both the pumping and intermediate states are strongly populated while the open state is depleted. At intermediate frequency, $f = 0.02$, the open state starts populating. In contrast, the pumping state starts depopulating, while the intermediate state follows likewise but most moderately. At high frequency, $f = 0.3 > f^*$, the pumping state is mostly unpopulated; the intermediate state exhibits a surviving signature and the open state is widely filled. The full evolution of the probability distribution profile with the pump frequency is shown in Fig. 2B. The pumping converts the two states model bound-open into a three states problem: one pumping state (bound), an intermediate state and an open state. We observe (i) an inversion of population from the pumping state to the intermediate state and (ii) that the open state is mostly populated from the intermediate state.

We investigate this inversion of population by examining the mechanism controlling the detaching events. We start by measuring the mean escaping times, $T_{b \to o}^{chain}$, $T_{b \to o}^{Brownian}$, respectively, for the simulated chain and Brownian systems. We observe in Fig 3A that the mean escaping time decreases rapidly with the oscillation frequency above a well-identified crossover. Furthermore, for a fixed interwall distance, the mean escaping time decreases with the wall oscillation amplitude. Finally, the mean escaping time increases with the dissipation coefficient, $\gamma$. These three observations hold for both the chain system and the Brownian particles. We rationalize these combined effects by considering, first, the Brownian case. We observe in Fig. 3B that all data collapse onto a single master curve. Thus, the theoretical predictions (Eq. 1) and (Eq. 2) are in very good agreement with the numerical results for low and moderate angular frequency. We use $\sigma = 0.4\ a$ for the effective trap potential which is measured from numerical simulations (see Fig. S2). The deviation at high angular frequency is expected as inertial terms which start playing a role in this frequency range, are neglected in the theoretical description.

Having rationalized the escaping times, we focus now on the open phases. When the ligand escapes from a receptor well it diffuses until it rebinds to a receptor. This receptor may be different from the departing one. We measure the open time defined as the mean time to switch from the open to the bound state, $T_{o \to b}^{chain}$ and $T_{o \to b}^{Brownian}$, for the chain system and the Brownian particle, respectively. We observe in Fig. 3C that the open times vary very weakly with the wall oscillation frequency and are primarily prescribed by the interwall distance $h_0$ (See also Fig. S6). As for the chain system, and given the error bars, the open times are approximatively independent on frequency. For $h_0 = 18$, the chain must extend away from its natural end-to-end distance ($R_{ee} = 15.7$), an energetical cost of entropic nature which increases the open time. In contrast, for $h_0 = 8$, the chain is compressed and the open time is significantly reduced. For the case of a Brownian particle, the errors are less pronounced and we observe for all parameter configurations, a plateau at low frequency followed by a slow decrease of the open times.

Finally, we observe in Fig. 3D an increase of the flux ratio $k_{b \to o}/k_{o \to b}$ with the receptor wall frequency for the chain and Brownian systems. At low frequency, the flux ratio converges towards the Kramers' two states model limit, as expected. At high frequency, the flux ratio strongly disagrees with Kramers' limit. Specifically, for strong binding, the flux ratio is strongly unbalanced towards unbinding events and exceeds by more than two orders of magnitude the Kramers' limit. At equilibrium, the transition from the bound states to the open states is reaction-controlled as predicted by Kramers' two state model (see supplementary information and Figure

S3-S6). In contrast at high frequency, the transition rate from the bound state to open state is strongly biased while the transition from open to bound states is mainly unaffected by the wall pumping. Their combined effect populates efficiently the open state. These two regimes are in perfect agreement with our theoretical predictions.

Conclusion and discussion We introduce a protocol called mechanical pumping and demonstrated its effectiveness to prevent a ligand from binding to a receptor above a crossover frequency. By simulating both a Brownian particle and ligand attached to the end of a grafted polymer, we showed that the protocol does not depend on the particular details of the system. We show that the mechanism is driven by the relative friction between the monomers and the surrounding solvent with a mean escaping time depending exponentially on the monomer damping coefficient.

In this article, the Brownian case is used as a guidance while the polymeric system is more suitable for applications. However, for the latter case, several aspects remain to be discussed to assess its practical relevance. Firstly, the solvent is modelled as being implicit, which means that the molecular dynamics of the surrounding solvents is not simulated *per se*. However, the characteristic relaxation time of the solvent can be considered longer than the relevant range of oscillating period $1/f$. Thus, at short time scale, specifically during a couple of periods of oscillations, the pulled chain and the melt have a net relative motion and thus, friction. Additionally, in both theory and simulations, we neglected any dynamical hydrodynamic effects of the solvent as effects, like *e.g.,* acoustic streaming, are expected to arise at much longer time scale. Nevertheless, and while being beyond the scope of this article, the deterministic interactions between a soft or rigid wall and a floating object in Stokes regimes are usually known to generate a lifting, i.e. repulsive, force (56,57) over a large range of physical parameters. This additional effect is expected to enhance the mechanical pumping.

Secondly, in practice, the system can be implemented as sketched in Fig1A, wherein the receptor is attached onto a colloidal particle which can be set into motion by, e.g., and acoustic wave or an optical trapping. We may pursue one step ahead and design supramolecular networks, i.e., a gel, as pictured in Fig 4A&C made of ligands at the end of polymer chains, receptors and two types of colloidal particles. The polymer chains are grafted to the smallest particles. In contrast, the receptors are grafted onto the largest particles. As the speed of displacement upon radiation pressure scales as the particle size, this configuration ensures that the receptors are co-displaced with the largest particles, while the chain with floppy backbones are attached to particles which are weakly displaced. Interestingly, below the crossover frequency, the system is expected to behave as a soft solid and becomes liquid-like beyond. We note that this concept may be leveraged in mechanochemistry to induce liquid-gel transition with a mechanism differing from ultrasound-induced cavitation (44-47).

Thirdly and finally, we may wonder how realistic the crossover frequency is for practical applications and its dependency on the temperature. As the $f^*$ depends on the inverse of the monomeric friction coefficient $\gamma$, its evolution with temperature can be estimated by re-expressing the empirical Williams-Landel-Ferry (WLF) time-temperature relation (58) as

$$\mathrm{Log}_{10}\left(\frac{f^*}{f^*_{\mathrm{WLF}}}\right) = \frac{C_1(T-T_g)}{C_2+(T-T_g)}, \quad (3)$$

where $C_1 = 17.44 \text{ K}^{-1}$ and $C_2 = 51.6 \text{ K}$ are two universal empirical coefficients. We denote $T_g$ the glass temperature and $f^*_{\mathrm{WLF}}$ (see supplementary information) an empirical constant which are both monomer-dependent. Eq.3 predicts the temperature dependence of the crossover frequency in a master curve as indicated in Fig. 4B. As we decrease the temperature towards the glass temperature, $f^*$ strongly decreases. As a consequence, even if $f^*$ may be large at room temperature, it will necessarily fall into an experimentally-accessible frequency range by decreasing the temperature above the glass transition. Following the WLF approach, and the crossover frequency dependence from our theory (see section Theoretical predictions), we conclude by providing some order of magnitude for $f^*$ for some commercial polymer systems, specifically for the polybutadiene (cis-1,4- and trans-1,4-polybutadiene, PB), the Polydimethylsiloxane (PDMS) and the styrene-butadiene (S-SBR) and for adhesion energy typical of a hydrogen bond (see Fig 4D and supplementary information). For low temperature, but above the glass transition, $f^*$ falls into the kHz-MHz regime which is accessible to standard piezoelectric devices, or Atomic Force Microscopy tips. At room temperature, the frequency range greatly exceeds ∼100 MHz. The latter frequency range is mostly beyond the reach of standard ultrasonic or piezoelectric devices. However, a flurry of ingenious optomechanical designs operating up to the 1GHz-100GHz mechanical regime (59,60) have emerged over the past decade. Combining mechanochemistry with optomechanics will allow experimental implementations of optomechanical chemical switches at room temperature and will open a new pathway toward the control of ligand-receptor, molecular, and supramolecular binding.

60. Gil-Santos, E., Labousse, M., Baker, C., Goetschy, A., Hease, W., Gomez, C., Lemaître, A., Leo, G., Ciuti, C. & Favero, I. Light-mediated cascaded locking of multiple nano-optomechanical oscillators. *Phys. Rev. Lett.* **118**, 063605 (2017).

**Acknowledgments**: ML would like to thank Joshua McGraw for insightful discussions about lifting forces between small objects and soft substrates. MH thanks Prof. Hartmut Löwen for his hospitality during the summer 2020 visit.

Funding: This work was funded by the ESPCI Paris and CNRS.

Author contributions:

    Conceptualization: LL & ML

    Methodology: MH, TG, LL ML

    Investigation: MH, TG, LL ML

    Visualization: MH

    Funding acquisition: LL

    Project administration: LL ML

    Supervision: LL ML

    Writing – original draft: ML

    Writing – review & editing: MH ML LL

Competing interests: Authors declare that they have no competing interests.

Data and materials availability: All data, code, and materials used in the analysis are available to any researcher for purposes of reproducing or extending the analysis upon reasonable request.

## Chemical equilibrium

## Mechanical pumping

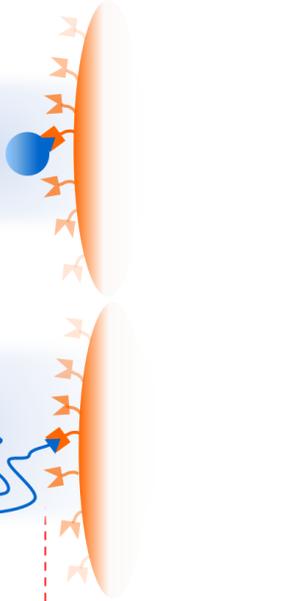

**A**

**C**

**E**

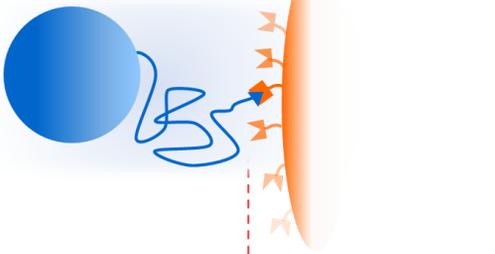

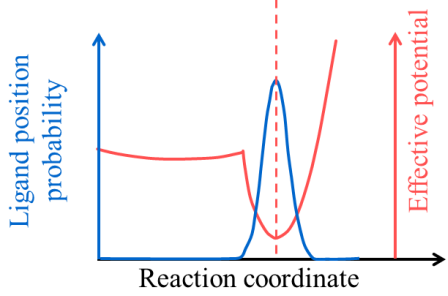

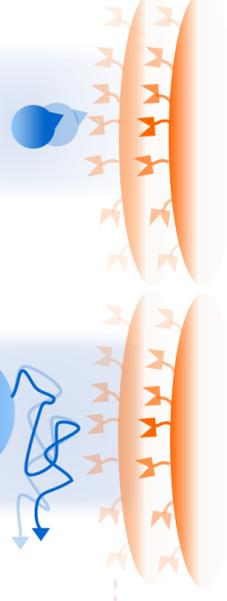

**B**

**D**

**F**

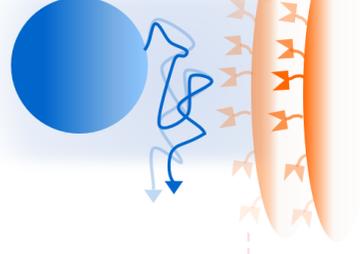

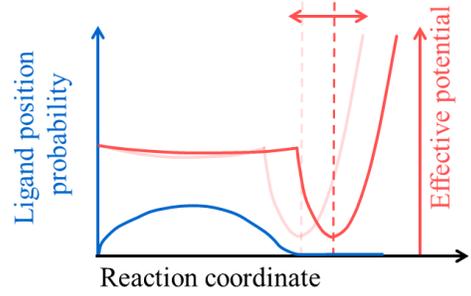

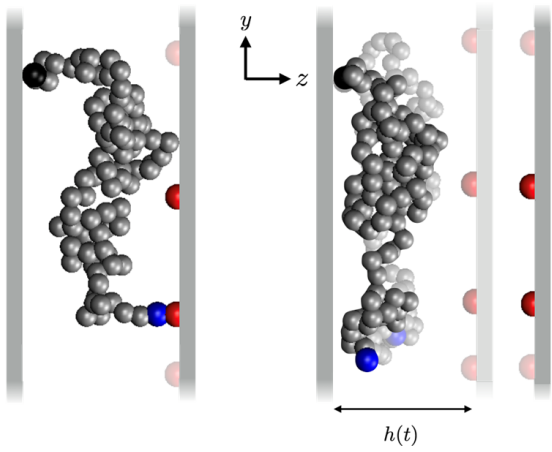

**G**

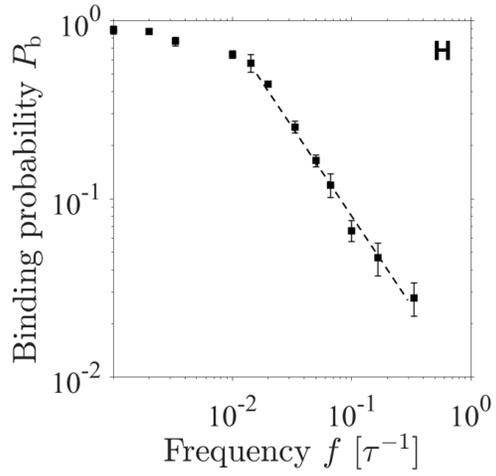

**H**

**Fig. 1. Controlling a reaction by mechanical pumping. (A)-(D)**. Sketch of the two systems with or without mechanical pumping **(A) & (B)** A Brownian particle may bind to a receptor positioned on the surface of a large particle which is either immobile (A) or set into oscillatory motion (B). **(C) & (D)** A ligand attached at the end of polymer grafted on a particle may bind to a receptor positioned on the surface of a large particle which is either immobile (C) or set into oscillatory motion (D). **(E)-(F)** Sketch of the principle of the chemical equilibrium (E) and during the mechanical pumping (F). (E) The ligand position density probability (blue line) is peaked in the middle of the receptor potential (red line). (F) The wall oscillates at an angular frequency $\omega$. (blue line) The ligand position density probability is widely spread outside the receptor positions. **(G)** Side view of a snapshot of the molecular dynamics simulation when the wall is static with (left) an interwall distance $h_0 = 8a$ or (right) when the wall moves sinusoidally around $h_0 = 8a$ with an amplitude $A = 2$ and a frequency $f = \omega/2\pi > f^*$. **(H)** Evolution of the ligand-receptor binding probability with the wall frequency for $h_0 = 8a$ and $A = 2$. The dash line represents a -1 slope as a guide for the eye. Error bars in (H) are obtained from 4 independent runs.

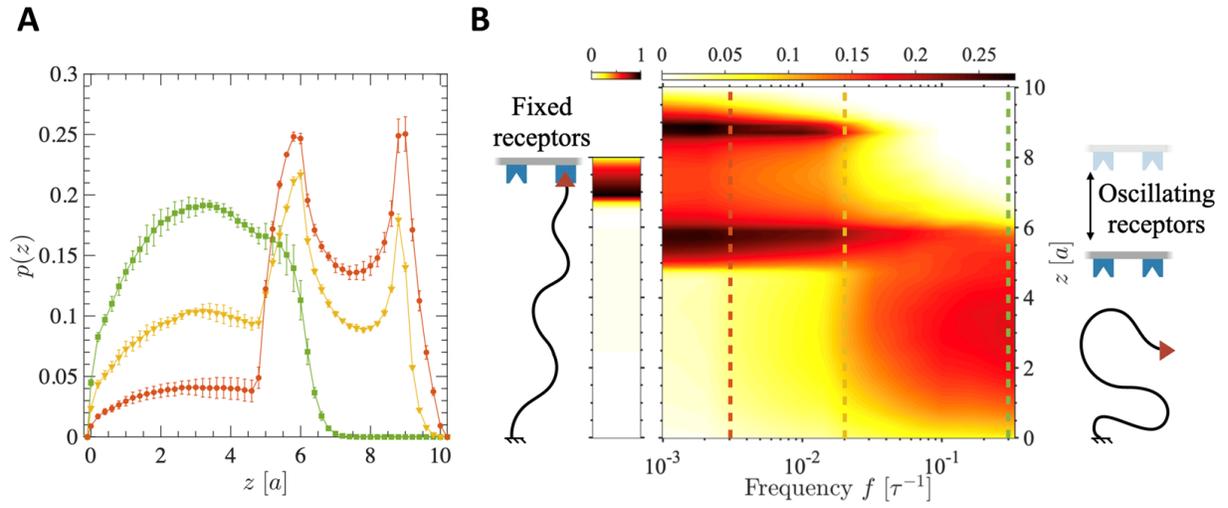

**Fig. 2 Mechanism of Gap opening**. **(A)** Ligand z-position probability density for $A = 2$, $h_0 = 8$ $\gamma = 10$ and $f = 3.3 \times 10^{-3}$ (dark orange circles), $f = 2.\times 10^{-2}$ (yellow triangles), $f = 3 \times 10^{-1}$ (green square). **(B)** Ligand z-position probability density for increasing frequency for $A = 2$, $h_0 = 8$ $\gamma = 10$. The three vertical dashed lines correspond to the situation shown in A. The left rectangular panel indicates the case of a static wall.

.

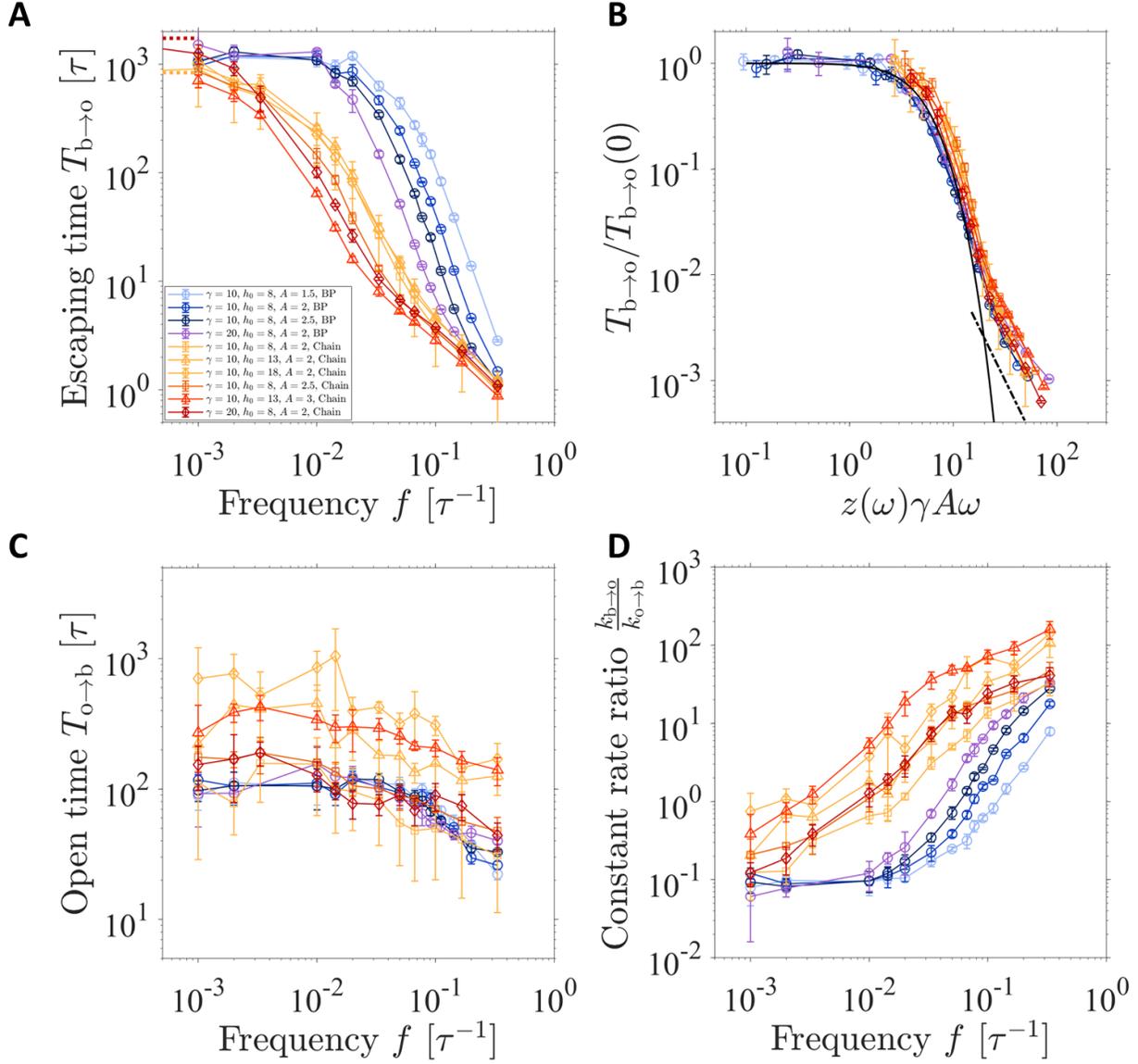

**Fig. 3. Binding and unbinding mechanisms and deviation from Kramers' theory**. The color code is common to panels A-D. Brownian particle, $h_0 = 8$, $\gamma = 10$, $A = 1.5, 2, 2.5$ (from light to dark blue). Brownian particle, $h_0 = 8$, $\gamma = 20$, $A = 2$ (purple circle). Chain, $\gamma = 10$ $A = 2$, $h_0 = 8, 13, 18$ (yellow square, yellow triangle, yellow diamond). Chain, $\gamma = 10$ $A = 2.5$, $h_0 = 8$ (orange square). Chain, $\gamma = 10$ $A = 3$, $h_0 = 13$ (dark-orange square). Chain, $\gamma = 20$ $A = 2$, $h_0 = 8$ (Bordeaux diamond). **(A)** Measure of the mean escaping time $T_{b \to o}$ as a function of the frequency. **(B)** Rationalization of the evolution of the dimensionless mean escaping time

$T_{b\to o}/T_{b\to o}(0)$ as a function of $z\gamma A\omega$ with $\omega = 2\pi f$ the angular frequency. $z = 1$ and $z = \sqrt{3\pi/2}(f\gamma)^{-1/2}$ for Brownian Particle and polymer system, respectively. **(C)** Measure of the open time $T_{o\to b}$ as a function of the frequency. **(D)** Evolution of the flux ratio $T_{o\to b}/T_{b\to o}$ with the wall frequency.

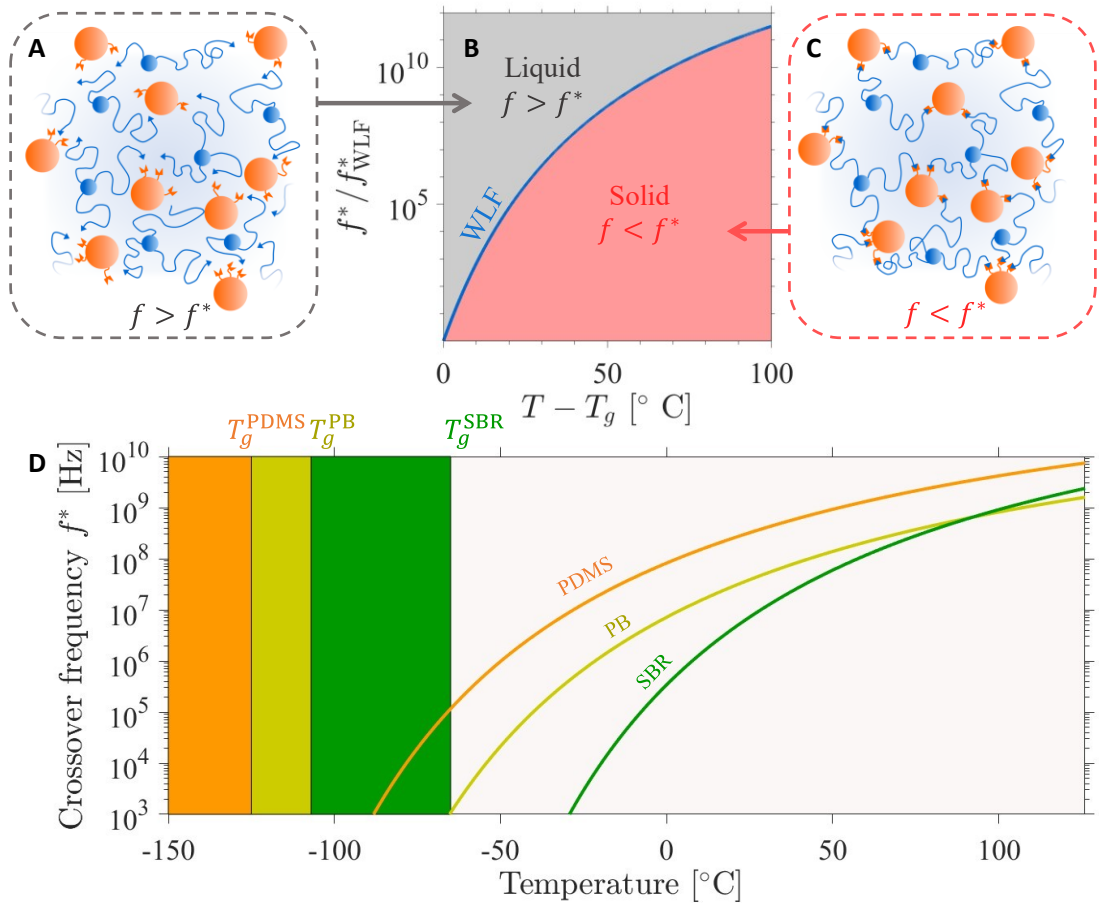

**Fig. 4 Evolution of the crossover frequency and prediction of liquid-solid transition in reticulated materials (A) & (C)** Schematics of a reticulated network above and below the unbinding crossover frequency. **(B)** Evolution of the normalized crossover frequency $f^*$ with the temperature above the glass transition. The transition is predicted from the WLF theory (blue line). Pink (grey, resp.) area denotes the region of the solid (liquid, resp.) phase. **(D)** Theoretical Predictions from WLF theory of the temperature dependence of the crossover frequency for three different polymers above their respective glass transition: PDMS (orange line), polybutadiene (anise green), styrene-butadiene (green).